\newcommand{\be}{\begin{equation}}\newcommand{\ee}{\end{equation}}
\newcommand{\bea}{\begin{eqnarray}}\newcommand{\eea}{\end{eqnarray}}
\newcommand{\nn}{\nonumber}\newcommand{\p}[1]{(\ref{#1})}
 \newcommand{\lb}[1]{\label{#1}}
 \newcommand\q{\quad}
\newcommand\qq{\quad\quad}
\newcommand\s{\scriptscriptstyle}

\def\a{\alpha}

\def\b{{\beta}}

\def\g{\gamma}

\def\d{\delta}

\def\eps{\epsilon}

\def\ve{\varepsilon}

\def\l{\lambda}
 \def\th{\theta}  \def\bt{\bar\theta}

\def\z{\zeta}

\def\L{\Lambda}

\def\pa{\partial}

\def\na{\nabla}

\newcommand\Tr{\mbox{Tr}\,}

\newcommand\ab{{\alpha\beta}}

\newcommand\A{{\s A}}

\newcommand{\0}{{\s 0}}

\newcommand{\pp}{{++}}
\newcommand{\m}{{--}}

\newcommand\cD{{\cal D}}

\def\sfrac#1#2{{\textstyle\frac{#1}{#2}}}

\documentclass[12pt]{article}
\usepackage{amsfonts}
\begin{document}

{\bf    SUPERSYMMETRIC CHERN-SIMONS MODELS IN HARMONIC SUPERSPACES}\\
\vspace{0.5cm}

{\bf B.M. Zupnik}\\

 {\it
Bogoliubov Laboratory of Theoretical Physics, JINR, Dubna,
  Moscow Region, 141980, Russia; E-mail: zupnik@theor.jinr.ru}

\begin{abstract}
We review harmonic superspaces of the $D{=}3, N{=}3$ and 4 supersymmetries
and gauge models in these superspaces. Superspaces of the $D{=}3, N{=}5$
supersymmetry use harmonic coordinates of the $SO(5)$ group.  The
superfield $N{=}5$ actions describe the off-shell
infinite-dimensional Chern-Simons supermultiplet.
\end{abstract}

\section{Introduction}

Supersymmetric extensions of the $D{=}3$ Chern-Simons theory were discussed in
 \cite{Si}-\cite{Schw}.
A superfield action of the $D{=}3, N{=}1$ Chern-Simons theory can be interpreted
as the superspace integral of the differential Chern-Simons superform
$dA+\sfrac{2i}3A^3$ in the framework of our theory of superfield integral
forms \cite{ZP1}-\cite{Z6}.

The Abelian $N{=}2$ CS action was first constructed in the $D{=}3,
N{=}1$ superspace \cite{Si}. The corresponding non-Abelian action was considered
in the $D{=}3, N{=}2$ superspace with the help of the Hermitian superfield
$V(x^m,\th^\a,\bt^\a)$, where $\th^\a$ and $\bt^\a$ are the complex conjugated
spinor coordinates \cite{ZP1}.  The unusual dualized form of the $N{=}2$ CS
Lagrangian contains the second vector field instead of the
scalar field\cite{NG}.

The $D{=}3, N{=}3$ CS theory was first analyzed by the
harmonic-superspace method \cite{ZK,Z3}.
 Supersymmetric action of the
$D{=}3, N{=}4$ Yang-Mills theory can also be  constructed in the
$D{=}3, N{=}3$ superspace, but the alternative formalism exists in
the $N{=}4$ superspace \cite{Z5}.

The authors of  \cite{HL} propose  using the $SO(5)/U(2)$ harmonic
superspace for the superfield description of the $D{=}3, N{=}5$ Chern-Simons
theory. The detailed analysis
of the superfield formalism of the Chern-Simons theory in this harmonic superspace
was presented in our recent paper \cite{Z9}. The alternative formalism of this theory using the $SO(5)/U(1)\times U(1)$
harmonics and additional harmonic conditions was considered in \cite{Z8}. It was
shown that the action of this model is invariant with respect to the
$D{=}3, N{=}6$ superconformal group. The superfield action without harmonic
constraints describes additional matter fields \cite{Z7}.

\setcounter{equation}0
\section{$N=4$ and $N=3$ harmonic superspaces}

We consider the following coordinates of the $D=3, N=4$ superspace:
\bea
&&z=(x^m, \th^\a_{i\hat{k}}),
\eea
where $i$ and $\hat{k}$ are two-component  indices of the automorphism groups
$SU_L(2)$ and $SU_R(2)$, respectively, $\a$ is the two-component index
of the $SL(2,R)$ group  and $m=0, 1, 2$ is the 3D vector index.
The $N=4$ supersymmetry transformations are
\bea
&&\d x^m=-i(\g^m)_\ab(\eps^\a_{j\hat{k}}\th^{\b j\hat{k}}-i\eps^\b_{j\hat{k}}
\th^{\a j\hat{k}}),
\eea
where $\g^m$ are the 3D $\gamma$ matrices.

The $SU_L(2)/U(1)$ harmonics $u^\pm_i$ \cite{GIKOS1} can be used to construct
the left analytic superspace \cite{Z5} with the $LA$ coordinates
\be
\z_L=(x^m_L, \th^{+\hat{k}\a}).
\ee
The $L$-analytic prepotential $V^{++}(\z_L,u)$ describes the left $N=4$
vector multiplet $A_m, \phi_{\hat{k}\hat{l}}, \l^\a_{i\hat{k}}, D^{ik}$.
The $D=3, N=4$ $SYM$ action can be constructed in terms of this prepotential
by analogy with the $D=4, N=2$ $SYM$ action \cite{Z1}.

Let us introduce the new notation for the left harmonics
$u_i^\pm=u^{(\pm1,0)}_i$ and the
analogous notation $v_{\hat{k}}^{(0,\pm1)}$ for the right $SL_R(2)/U(1)$
harmonics. The biharmonic $N=4$ superspace uses the Grassmann coordinates
\cite{Z5}
\be
\th^{(\pm1,\pm1)\a}=u^{(\pm1,0)}_iv_{\hat{k}}^{(0,\pm1)}\th^{i\hat{k}\a}.
\ee
In this representation, we have
\bea
&&\z_L=(x^m_L,\q \th^{(1,\pm1)\a}), \q V^{++}\equiv V^{(2,0)},\\
&&D^{(1,\pm1)}_\a V^{(2,0)}=0,\q D^{(0,2)}_vV^{(2,0)}=0.
\eea

The right analytic $N=4$ coordinates are
\bea
&&\z_R=(x^m_R, \q \th^{(\pm1,1)\a}),\q x^m_R=x^m_L-2i(\g^m)_\ab\th^{(-1,1)\a}
\th^{(1,1)\b}.
\eea
The mirror $R$ analytic prepotential $\hat{V}^{(0,2)}$
\bea
D^{(\pm1,1)}\hat{V}^{(0,2)}=0,\q D^{(2,0)}_u\hat{V}^{(0,2)}=0
\eea
describes the right
$N=4$ vector multiplet $\hat{A}_m, \hat\phi_{ij}, \hat\l^\a_{i\hat{k}},
\hat{D}_{ik}$, where $\hat{A}_m$ is the mirror vector gauge field
using the independent gauge group. The right $N=4$ $SYM$ action is similar
to the analogous left action. These multiplets can be formally connected
by the map $SU_L(2)\leftrightarrow SU_R(2)$.

The $N=4$ superfield  Chern-Simons type (or $BF$-type) action for the gauge group
$U(1)\times U(1)$
connects two mirror vector multiplets
\bea
&&\int du d^3x_L d\th^{(-4,0)}V^{(2,0)}(\z_L,u)D^{(1,1)\a}D^{(1,1)}_\a
\hat{V}^{(0,-2)},
\eea
where the right connection satisfies the equation
\bea
\cD^{(0,2)}_v\hat{V}^{(0,-2)}=\cD^{(0,-2)}_v\tilde{V}^{(0,2)},\q
\cD^{(2,0)}_u\hat{V}^{(0,-2)}=0.
\eea
The component form of this action was considered in \cite{BG,KS}.

We can identify the left and right isospinor indices in the $N=4$
spinor coordinates
\bea
\th^\a_{j\hat{k}}~\rightarrow~\th^\a_{jk}=\th^\a_{(jk)}+\sfrac12\ve_{jk}\th^\a,
\eea
where the Grassmann coordinates $\th^\a_{(jk)}$ describe the $N=3$ superspace.
The harmonic superspace of the $D=3, N=3$ supersymmetry uses the standard
harmonics $u^\pm_i$\cite{ZK,Z3}
\bea
&&x^\ab_{3}=x^\ab
+i(\theta^{\alpha\pp}\theta^{\beta\m}+\theta^{\beta\pp}\theta^{\alpha\m})~,
\lb{C1}\\
&&\theta^{++\alpha}=u^+_iu^+_k\th^{(ik)\a}~,
\quad\theta^{--\alpha}=u^-_iu^-_k\th^{(ik)\a}~,
\quad \theta^{\alpha\0}=u^+_iu^-_k\th^{(ik)\a}
~,\lb{C2}\\
&&\d x^\ab_{3}=-2i\eps^{--\a}\th^{++\b}-2i\eps^{--\b}\th^{++\a}+
2i\eps^{0\a}\th^{0\b}.\nn
\eea

The vector $N=3$ supermultiplet is described
by the analytic superfield $V^{++}(x_3, \th^{++}, \th^0, u)$ and the
corresponding analytic $N=3$ superfield strength is
\be
W^{++}(x_3, \th^{++}, \th^0, u)=-\sfrac12D^{++\a}D^{++}_\a V^{--}.
\ee
The action of the corresponding
$CS$-theory can be constructed in the full or analytic $N=3$ superspaces
\cite{ZK,Z3}.

\setcounter{equation}0
\section{Harmonic superspaces for the group $SO(5)$}
The homogeneous space $SO(5)/U(2)$ is parametrized by elements of the harmonic
 5$\times$5 matrix
\be
U^K_a=(U^{+i}_a, U^0_a, U^-_{ia})
=(U^{+1}_a, U^{+2}_a, U^0_a, U^-_{1a}, U^-_{2a}),
\ee
where $a=1, \ldots 5$ is the vector index of the group $SO(5)$, $i=1, 2$ is
the spinor index of the group $SU(2)$, and $U(1)$-charges are denoted by symbols
$+, -, 0$. The basic relations for these harmonics are
\bea
&&U^{+i}_aU^{+k}_a=U^{+i}_aU^0_a=0,\q U^-_{ia}U^-_{ka}=U^-_{ia}
U^0_a=0,\q U^{+i}_aU^-_{ka}=
\d^i_k,\q U^0_aU^0_a=1,\nn\\
&&U^{+i}_aU^-_{ib}+U^-_{ia}U^{+i}_b+U^0_aU^0_b=\d_{ab}.
\eea

We consider the $SO(5)$ invariant harmonic derivatives with nonzero
$U(1)$ charges
\bea
&&\pa^{+i}=U^{+i}_a\frac{\pa}{\pa U^0_a}-U^0_a\frac{\pa}{\pa U^-_{ia}},\q
\pa^{+i}U^0_a=U^{+i}_a,\q \pa^{+i}U^-_{ka}=-\d^i_kU^0_a,\nn\\
&&\pa^{++}=U^+_{ia}\frac{\pa}{\pa U^-_{ia}},\q [\pa^{+i},\pa^{+k}]=
\ve^{ki}\pa^{++},\q\pa^{+i}\pa^{+}_i=\pa^{++},\nn\\
&&\pa_i^-=U^-_{ia}\frac{\pa}{\pa U^0_a}-U^0_a\frac{\pa}{\pa U^{+i}_a},
\q \pa_i^-U^0_a=U^-_{ia},\q \pa_i^-U^{+k}_a=-\d^k_iU^0_a,\\
&&\pa^{--}=U^{-i}_{a}\frac{\pa}{\pa U^{+i}_{a}},\q[\pa_i^-,\pa_k^-]=
\ve_{ki}\pa^{--},\q \pa^{-k}\pa_k^-=-\pa^{--},\nn
\eea
where some relations between these harmonic derivatives are defined.
The $U(1)$ neutral harmonic derivatives form the Lie algebra $U(2)$
\bea
&&\pa^i_k=U^{+i}_a\frac{\pa}{\pa U^{+k}_{a}}-U^{-}_{ka}\frac{\pa}
{\pa U^-_{ia}},\q[\pa^{+i},\pa^-_k]=-\pa^i_k,\\
&&\pa^0\equiv \pa^k_k=U^{+k}_a\frac{\pa}{\pa U^{+k}_{a}}-U^{-}_{ka}
\frac{\pa}{\pa U^-_{ka}},\q [\pa^{++},\pa^{--}]=\pa^0,\nn\\
&&\pa^i_k U^{+l}_a=\d_k^lU^{+i}_a,\q \pa^i_k U^-_{la}=-\d^i_lU^-_{ka}.
\eea
The operators $\pa^{+k}, \pa^{++}, \pa^-_k, \pa^{--}$ and $\pa^i_k$ satisfy
the commutation relations of the Lie algebra $SO(5)$.

One defines an ordinary complex conjugation on these harmonics
\bea
&&\overline{U^{+i}_{a}}=U^-_{ia},\q \overline{U^{0}_{a}}=U^{0}_{a},
\eea
however, it is convenient to use a special conjugation in the harmonic
space
\bea
&&(U^{+i}_a)^\sim=U^{+i}_a,\q (U^-_{ia})^\sim=U^-_{ia},\q
(U^{0}_{a})^\sim=U^{0}_{a}.
\eea
All harmonics are real with respect to this conjugation.

The full superspace of the $D{=}3, N{=}5$ supersymmetry has the spinor
$CB$ coordinates $\th^\a_a,\q (\a=1, 2;\q a=1, 2, 3, 4, 5)$ in
addition to the coordinates $x^m$ of the three-dimensional Minkowski
space. The group $SL(2,R)\times SO(5)$ acts on the spinor coordinates.
The superconformal transformations of these coordinates are considered in
Appendix.

The $SO(5)/U(2)$ harmonics allow us to construct projections of the spinor
coordinates and the partial spinor derivatives
\bea
&&\th^{+i\a}=U^{+i}_a\th^\a_a,\q \th^{0\a}=U^0_a\th^\a_a,\q
\th^{-\a}_i=U^-_{ia}\th^\a_a,\\
&&\pa^-_{i\a}=\pa/\pa\th^{+i\a},\q \pa^0_\a=\pa/\pa\th^{0\a},\q
\pa^{+i}_\a=\pa/\pa\th^{-\a}_i.\nn
\eea

The analytic coordinates ($AB$-representation) in the full harmonic superspace
use these projections of 10 spinor coordinates
$\th^{+i\a}, \th^{0\a}, \th^{-\a}_i$ and the following representation
of the vector coordinate:
\bea
&&x^m_\A\equiv y^m=x^m+i(\th^{+k}\g^m\th^-_k)=x^m+i(\th_a\g^m\th_b)
 U^{+k}_aU^-_{kb}.
\eea

The analytic coordinates are real with respect to the special conjugation.

The harmonic derivatives have the following form in $AB$:
\bea
&&\cD^{+k}=\pa^{+k}-i(\th^{+k}\g^m\th^{0})\pa_m+\th^{+k\a}\pa^0_\a
-\th^{0\a}\pa^{+k}_\a,\nn\\
&&\cD^{++}=\pa^{++}+i(\th^{+k}\g^m\th^+_k)\pa_m+\th^{+\a}_k\pa^{+k}_{\a}\lb{ABharm},\\
&&\cD^{k}_l=\pa^k_l+\th^{+k\a}\pa^-_{l\a}-\th^{-\a}_l\pa^{-k}_{\a}.\nn
\eea

We use the commutation relations
\bea
&&[\cD^{+k},\cD^{+l}]=-\ve^{kl}\cD^{++},\q \cD^{+k}\cD^+_k=\cD^{++}.
\eea

The AB spinor derivatives are
\bea
&&D^{+i}_\a=\pa^{+i}_\a,\q D^-_{i\a}=-\pa^-_{i\a}-2i\th^{-\b}_i\pa_\ab,
\nn\\
&&D^0_\a=\pa^0_\a+i\th^{0\b}\pa_\ab.
\eea

The coordinates of the analytic superspace $\z=(y^m, \th^{+i\a},
\th^{0\a},U_a^K)$ have the Grassmann dimension  6 and  dimension of the
even space  3+6. The functions  $\Phi(\z)$ satisfy the Grassmann analyticity
condition in this superspace
\be
D^{+k}_\a\Phi=0.
\ee
In addition to this condition, the analytic superfields in the
$SO(5)/U(2)$ harmonic superspace possess also the $U(2)$-covariance.
This subsidiary condition looks especially simple for the  $U(2)$-scalar
superfields
\be
\cD^k_l\L(\z)=0.\lb{U2a}
\ee

The integration measure in the analytic superspace $d\mu^{(-4)}$ has
dimension zero
\bea
&&d\mu^{(-4)}=dUd^3x_\A(\pa^0_\a)^2(\pa^-_{i\a})^4=dUd^3x_\A d\th^{(-4)}.
\lb{meanal}
\eea

The $SO(5)/U(1)\times U(1)$  harmonics can be
defined via the components of the real orthogonal 5$\times$5
matrix \cite{Z8,Z7}
\be
U^K_a=\left(U^{(1,1)}_a, U^{(1,-1)}_a, U^{(0,0)}_a,
U^{(-1,1)}_a, U^{(-1,-1)}_a\right)
\ee
where $a$ is the SO(5) vector index and the index $K=1, 2,\ldots 5$
corresponds to given combinations of the U(1)$\times$U(1) charges.

We  use the following harmonic derivatives
\bea
&&\pa^{(2,0)}=U^{(1,1)}_b\pa/\pa
U^{(-1,1)}_b-U^{(1,-1)}_b\pa/\pa U^{(-1,-1)}_b,\nn\\
&&\pa^{(1,1)}=U^{(1,1)}_b\pa/\pa
U^{(0,0)}_b-U^{(0,0)}_b\pa/\pa U^{(-1,-1)}_b,\nn\\
&&\pa^{(1,-1)}=U^{(1,-1)}_b\pa/\pa
U^{(0,0)}_b-U^{(0,0)}_b\pa/\pa U^{(-1,1)}_b,\nn\\
&&\pa^{(0,2)}=U^{(1,1)}_b\pa/\pa
U^{(1,-1)}_b-U^{(-1,1)}_b\pa/\pa U^{(-1,-1)}_b,\nn\\
&&\pa^{(0,-2)}=U^{(1,-1)}_b\pa/\pa
U^{(1,1)}_b-U^{(-1,-1)}_b\pa/\pa U^{(-1,1)}_b.
\eea

We define the harmonic projections of the $N{=}5$ Grassmann
coordinates
\bea
&&\th^K_\a=\th_{a\a}U^K_a=(\th^{(1,1)}_\a, \th^{(1,-1)}_\a, \th^{(0,0)}_\a,
\th^{(-1,1)}_\a, \th^{(-1,-1)}_\a).
\eea
The $SO(5)/U(1)\times U(1)$ analytic superspace contains only  spinor coordinates
\bea
&&\zeta=( x^m_\A, \th^{(1,1)}_\a, \th^{(1,-1)}_\a, \th^{(0,0)}_\a),\\
&&x^m_\A=x^m+i\th^{(1,1)}\g^m\th^{(-1,-1)}+i\th^{(1,-1)}\g^m\th^{(-1,1)},\nn
\\
&&\d_\eps x^m_\A=-i\eps^{(0,0)}\g^m\th^{(0,0)}-2i\eps^{(-1,1)}\g^m\th^{(1,-1)}
-2i\eps^{(-1,-1)}\g^m\th^{(1,1)},
\eea
where $\eps^{K\a}=\eps^\a_a U^K_a$ are the harmonic projections of
the supersymmetry parameters.

General superfields in the analytic coordinates depend also on
additional spinor coordinates $\th^{(-1,1)}_\a$ and $\th^{(-1,-1)}_\a$.
 The harmonized partial spinor derivatives are
\bea
&&\pa^{(-1,-1)}_\a=\pa/\pa\th^{(1,1)\a},\q \pa_\a^{(-1,1)}=
\pa/\pa\th^{(1,-1)\a},\q\pa^{(0,0)}_\a=\pa/\pa\th^{(0,0)\a},\lb{partspin}\\
&&\pa^{(1,1)}_\a=\pa/\pa\th^{(-1,-1)\a},\q\pa^{(1,-1)}_\a=\pa/\pa
\th^{(-1,1)\a}.\nn
\eea

We use  the special conjugation $\sim~$ in the harmonic superspace
\bea
&&\widetilde{U^{(p,q)}_a}=U^{(p,-q)}_a,\q
\widetilde{\th^{(p,q)}_\a}=\th^{(p,-q)}_\a,
\q\widetilde{x^m_\A}=x^m_\A,\nn\\
&&(\th^{(p,q)}_\a\th^{(s,r)}_\b)^\sim=\th^{(s,-r)}_\b\th^{(p,-q)}_\a,\q
\widetilde{f(x_\A)}=\bar{f}(x_\A),
\eea
where $\bar{f}$ is the ordinary complex conjugation. The analytic superspace is
real with respect to the special conjugation.

The analytic-superspace integral measure contains partial
spinor derivatives \p{partspin}
\bea
&&d\mu^{(-4,0)}=-\frac{1}{64}dU d^3x_\A
(\pa^{(-1,-1)})^2(\pa^{(-1,1)})^2(\pa^{(0,0)})^2=
dUd^3x_\A d^6\th^{(-4,0)},\lb{muanal}\\
&&\int  d^6\th^{(-4,0)}(\th^{(1,1)})^2(\th^{(1,-1)})^2(\th^{(0,0)})^2=1.
\nn
\eea

The harmonic derivatives of the analytic basis commute with the
generators of the $N{=}5$ supersymmetry
\bea
&&\cD^{(1,1)} =\pa^{(1,1)}-i\th^{(1,1)}_\a\th^{(0,0)}_\b\pa^\ab-
\th^{(0,0)\a}\pa^{(1,1)}_\a+\th^{(1,1)\a}\pa^{(0,0)}_\a,\nn\\
&&\cD^{(1,-1)}=\pa^{(1,-1)}-i\th^{(1,-1)}_\a\th^{(0,0)}_\b\pa^\ab-
\th^{(0,0)\a}\pa^{(1,-1)}_\a+\th^{(1,-1)\a}\pa^{(0,0)}_\a=
-(\cD^{(1,1)})^\dagger,\nn\\
&&\cD^{(2,0)}=[\cD^{(1,-1)},\cD^{(1,1)}]
=\pa^{(2,0)}-2i\th^{(1,1)}_\a\th^{(1,-1)}_\b\pa^\ab-\th^{(1,-1)\a}\pa^{(1,1)}_\a
+\th^{(1,1)\a}\pa^{(1,-1)}_\a,\nn\\
&&\cD^{(0,2)}=\pa^{(0,2)}+\th^{(1,1)\a}\pa^{(-1,1)}_\a
-\th^{(-1,1)\a}\pa^{(1,1)}_\a\nn\\
&& \cD^{(0,-2)}=-(\cD^{(0,2)})^\dagger=\pa^{(-2,0)}
+\th^{(1,-1)\a}\pa^{(-1,-1)}_\a-\th^{(-1,-1)\a}\pa^{(1,-1)}_\a.\nn
\eea

It is useful to
define the AB-representation of the U(1) charge operators
\bea
&&\cD^0_1 A^{(p,q)}=p\,A^{(p,q)},\q \cD^0_2
A^{(p,q)}=q\,A^{(p,q)},
\eea
where $A^{(p,q)}$ is an arbitrary harmonic superfield in AB.

The spinor derivatives  in the analytic basis are
\bea
&&D^{(-1,-1)}_\a=\pa^{(-1,-1)}_\a+2i\th^{(-1,-1)\b}\pa_\ab,\q
D^{(-1,1)}_\a=\pa^{(-1,1)}_\a+2i\th^{(-1,1)\b}\pa_\ab,\nn\\
&& D^{(0,0)}_\a=\pa^{(0,0)}_\a+i\th^{(0,0)\b}\pa_\ab,\q
D^{(1,1)}_\a=\pa^{(1,1)}_\a,\q D^{(1,-1)}_\a=\pa^{(1,-1)}_\a.
\eea

\setcounter{equation}0
\section{$N=6$ Chern-Simons theory in
  harmonic superspaces}

The harmonic derivatives  $\cD^{+k}, \cD^{++}$ together with the spinor
derivatives $D^{+k}_\a$ determine the $CR$-structure of the harmonic
 $SO(5)/U(2)$ superspace. The $U(2)$-covariant $CR$-structure is
invariant with respect to the $N{=}5$ supersymmetry. This $CR$-structure
should be preserved in the superfield gauge theory.

The gauge superfields (prepotentials) $V^{+k}(\z)$ and $V^{++}(\z)$ in the
harmonic $SO(5)/U(2)$ superspace satisfy the following conditions of the
Grassmann analyticity and $U(2)$-covariance:
\bea
 D^{+k}_\a V^{+l}=D^{+k}_\a V^{++}=0,\q
\cD^i_jV^{+k}=\d^k_jV^{+i},\q \cD^i_jV^{++}=\d^i_jV^{++}.
\lb{U2V}
\eea
In the gauge group $SU(n)$, these traceless matrix superfields are anti-Hermitian
\be
(V^{+k})^\dagger=-V^{+k},\q (V^{++})^\dagger=-V^{++},
\ee
where operation $\dagger$ includes the transposition and $\sim$-conjugation.

Analytic superfield parameters of the gauge group $SU(n)$ satisfy the conditions
of the generalized $CR$ analyticity
\be
D^{+k}_\a\L=\cD^i_j\L=0,\lb{CR}
\ee
they are traceless and anti-Hermitian $\L^\dagger=-\L$.

We treat these prepotentials as connections in the covariant gauge derivatives
\bea
&&\na^{+i}=\cD^{+i}+V^{+i},\q \na^{++}=\cD^{++}+V^{++},\nn\\
&&\d_\L V^{+i}=\cD^{+i}\L +[\L,V^{+i}],\q
\d_\L V^{++}=\cD^{++}\L +[\L,V^{++}],\lb{Vtrans}\\
&&D^{+k}_\a\d_\L V^{+k}=D^{+k}_\a\d_\L V^{++}=0,\q
\cD^i_j\d_\L V^{+k}=\d^k_j\d_\L V^{+i},\q \cD^i_j\d_\L V^{++}
=\d^i_j\d_\L V^{++},\nn
\eea
where the infinitesimal gauge transformations of the gauge superfields are
defined. These covariant derivatives commute with the spinor derivatives
$D^{+k}_\a$ and preserve the $CR$-structure in the harmonic superspace.

We can construct three analytic superfield strengths off the mass shell
\bea
&&F^{++}=\sfrac12\ve_{ki}[\na^{+i},\na^{+k}]=V^{++}-\cD^{+k}V^{+}_k-
V^{+k}V^+_k,\nn\\
&&F^{(+3)k}=[\na^{++},\na^{+k}]=\cD^{++}V^{+k}-\cD^{+k}V^{++}+
[V^{++},V^{+k}].\lb{ancurv}\nn
\eea

The superfield action in the analytic $SO(5)/U(2)$ superspace is defined
on three prepotentials $V^{+k}$ and $V^{++}$ by analogy with the off-shell
action of the $SYM_4^3$ theory \cite{GIKOS2}
\bea
S_1=\frac{ik}{12\pi}\int d\mu^{(-4)}\Tr\{V^{+j}\cD^{++}V^+_j
+2V^{++}\cD^+_jV^{+j}+(V^{++})^2+V^{++}[V^+_j,V^{+j}]\},\lb{act1}
\eea
where $k$ is the coupling constant, and a choice of the numerical multiplier
guarantees the correct normalization of the vector-field action. This action
is invariant with respect to the infinitesimal gauge transformations of
the prepotentials \p{Vtrans}. The idea of construction of the superfield
action in the harmonic $SO(5)/U(2)$ was proposed in \cite{HL}, although
the detailed construction of the superfield Chern-Simons theory was
not discussed in this work. The equivalent superfield action was considered
in the framework of the alternative superfield formalism \cite{Z8}.
The superconformal $N=5$ invariance of this action was proven in \cite{Z9}.

The action $S_1$ yields superfield equations of motion which mean triviality
of the superfield strengths of the theory
\bea
&&F^{(+3)}_k=\cD^{++}V^+_k-\cD^+_kV^{++}+[V^{++},V^+_k]=0,\nn\\
&&F^{++}=V^{++}-\cD^{+k}V^{+}_k-V^{+k}V^+_k=0.\lb{equ}
\eea
These classical superfield equations have pure gauge solutions for the
prepotentials only
\be
V^{+k}=e^{-\L}\cD^{+k}e^\L,\q V^{++}=e^{-\L}\cD^{++}e^\L,
\ee
where $\L$ is an arbitrary analytic superfield.

The transformation of the sixth supersymmetry can be defined on the analytic
$N{=}5$ superfields
\bea
&&\d_6V^{++}=\eps^\a_6 D^0_\a V^{++},\q \d_6V^{+k}=\eps^\a_6 D^0_\a
V^{+k},\\
&&\d_6\cD^{+k}V^{+l}=\eps^\a_6 D^0_\a\cD^{+k}V^{+l},\q
\d_6\cD^{++}V^{+l}=\eps^\a_6 D^0_\a\cD^{++}V^{+l},\lb{6susy}
\eea
where $\eps^\a_6$ are the corresponding odd parameters. This transformation
preserves the Grassmann analyticity and $U(2)$-covariance
\be
\{D^0_\a,D^{+k}_\b\}=0,\q [\cD^k_l,D^0_\a]=0,\q [\cD^{+k},D^0_\a]
=D^{+k}_\a,\q [\cD^{++},D^0_\a]=0.
\ee
The action $S_1$ is invariant with respect to this sixth supersymmetry
\be
\d_6S_1=\int d\mu^{(-4)}\eps^\a_6 D^0_\a L^{(+4)}=0.
\ee

In the $SO(5)/U(1)\times U(1)$ harmonic superspace, we can introduce
the  $D{=}3, N{=}5$ analytic matrix gauge prepotentials
corresponding to the five harmonic derivatives
\bea
&&V^{(p,q)}(\z,U)=[V^{(1,1)},\q V^{(1,-1)},\q V^{(2,0)},\q V^{(0,\pm2)}],
\nn\\
&&(V^{(1,1)})^\dagger=-V^{(1,-1)},\q
(V^{(2,0)})^\dagger=V^{(2,0)},\q
V^{(0,-2)}=[V^{(0,2)}]^\dagger, \lb{real}
\eea
where the Hermitian conjugation $\dagger$ includes $\sim$ conjugation of
matrix elements and transposition.

We shall consider the restricted gauge supergroup using the
supersymmetry-preserving harmonic ($H$) analyticity
constraints on the gauge superfield parameters
\be
H1:\qq \cD^{(0,\pm2)}\L=0.\lb{LHA}
\ee
These constrains yield  additional reality conditions for the component gauge
parameters.

We use   the harmonic-analyticity constraints on
the gauge prepotentials
\be
H2:\qq V^{(0,\pm2)}=0,\q
\cD^{(0,-2)}V^{(1,1)}=V^{(1,-1)}, \q \cD^{(0,2)}V^{(1,1)}=0
\ee
and the conjugated constraints  combined with relations
\p{real}.

The superfield CS action
can be constructed in terms of these $H$-constrained gauge superfields
\cite{Z7}
\bea
&&S=-\frac{2ik}{12\pi }\int   d\mu^{(-4,0)}\Tr\{V^{2,0}\cD^{(1,-1)}
V^{(1,1)}+V^{1,1}\cD^{(2,0)}V^{(1,-1)}\nn\\
&&+V^{1,-1}\cD^{(1,1)}V^{(2,0)}+V^{2,0}[V^{(1,-1)},V^{(1,1)}]-\sfrac12
V^{(2,0)}V^{(2,0)}\}.\lb{CSact}
\eea

Note, that the similar harmonic superspace based on the $USp(4)/U(1)\times U(1)$
harmonics
was used in \cite{BLS} for the harmonic interpretation of the $D=4, N=4$
super Yang-Mills constraints.

This work was partially supported by
the grants RFBR 06-02-16684, DFG 436 RUS 113/669-3, INTAS 05-10000008-7928
and by the Heisenberg-Landau programme.

\end{document}